# Non-stoichiometric and Subnano-heterogeneous Ln-incorporated UO₂: its defect chemistry and thermal oxidation

Juejing Liu[1], Shinhyo Bang[2], Natalie S. Yaw[1], Sam Karcher[2], Emma C. Kindall[1], Arjen van Veelen[3], Steven Conradson[2], Nicolas Clavier[4], John McCloy[1,2], Nicolas Dacheux[4], Xiaofeng Guo[1,2,*]

[1] *School of Mechanical and Materials Engineering, Washington State University, Pullman, Washington 99164, United States*
[2] *Department of Chemistry, Washington State University, Pullman, Washington 99164, United States*
[3] *Materials Science and Technology Division, Los Alamos National Laboratory, Los Alamos, New Mexico 87545, United States*
[4] *ICSM, Univ Montpellier, CNRS, CEA, ENSCM, Site de Marcoule, Bagnols sur Cèze, 30207, France*

**Abstract**

The defect chemistry and thermal oxidation of lanthanide (Ln) incorporated-$UO_2$ are critical for understanding and predicting their behavior as enhanced fuels, mixed oxide (MOX) fuels, spent nuclear fuels (SNF), and particles for safeguard purposes. In this study, we independently controlled the Ln type ($Ce^{4+}$, $Nd^{3+}$, and $Gd^{3+}$) and the preparation condition (reduced and nonreduced) to investigate their correlations to the generated non-equilibrated defects correspondingly. From early to late lanthanides: Ce and U formed close-to-ideal solid solutions in *Fm-3m* and oxidized to $(Ce, U)_4O_9$, Nd and U mixing under the reducing condition formed solid solutions with oxygen vacancies ($V_o^{\cdot\cdot}$) aggregating near Nd, and the mixing of smaller Gd with U resulted in short-range subnano-domain segregations with *Ia-3* region embedded in the global *Fm-3m* matrix. Both trivalent Ln-incorporated $UO_2$ oxidized to a mixture of $(Ln, U)_4O_9$ and $(Ln, U)_3O_8$. From these signature defect structures resulting from both Ln type and preparation condition, we proposed kinetic model and thermodynamic hypothesis for explaining the oxidation resistance of $(Ln, U)O_2$. Although originated from f-block oxides, the discovery of long-range disorder short-range ordering may be not uncommon in other metal oxide systems, which can strongly influence their functionalities and properties.

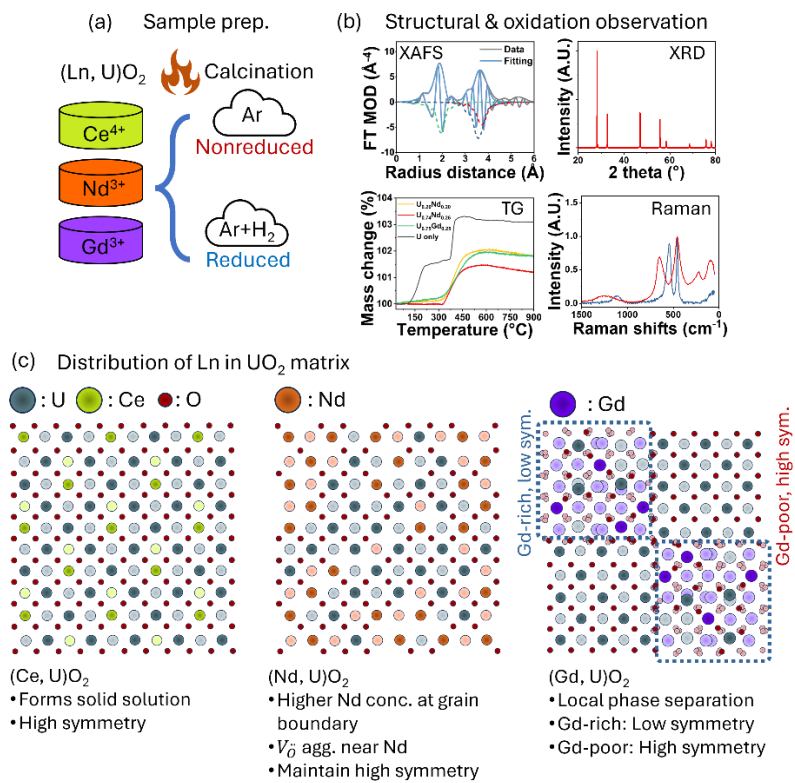

**Key words**: uranium dioxide, lanthanide, solid solution, defect chemistry, thermal oxidation, phase segregation, short-range ordering

**Introduction**

Nuclear power provides a reliable, weather-independent source of low-carbon electricity. To continue the development of nuclear energy, advancements in the safety and efficiency of nuclear fuel fabrication, use, recycling (e.g., mixed oxide fuel, MOX), and disposal are essential.[1-3] A safer fuel design benefits the entire fuel cycle and reactor operation.[4,5] Doping $UO_2$ with metals enhances properties of fuels. For instance, Cr, Zr, Mn, and Ti dopant provides following benefits:[5-9] preventing uranium oxidation, promoting larger grain size for fission gas retention, and enhancing thermal conductivity.

Lanthanide (Ln)-incorporated $UO_2$, or (Ln, U)$O_2$, opens opportunities for studying spent nuclear fuel (SNF) or MOX simulated fuel materials, as well as potentially improving fuel properties (e.g., oxidation resistance). As major fission products, Ln are readily incorporated by the $UO_2$ fluorite structure to form solid solutions. Although Ln doping generally decreases Young's modulus and thermal conductivity of $UO_2$,[6,10,11] it was found to inhibit $UO_2$ oxidation by preventing interstitial O ($O_i''$) formation.[10,12-15] Understanding Ln behavior within the $UO_2$ matrix also help design storage strategy as well as efficient separation techniques for recycling and reprocessing SNF.[16,17] Lastly, studying high concentration of Ln in $UO_2$ also simulates MOX fuel due to the analogs of Ln for Pu, where the fabricated MOX often show cation-charge distribution and phase segregation of U- vs Pu-rich oxide regions.[18,19]

Previously, the defect chemistry of (Ln, U)$O_2$ has been studied extensively.[3,5,10,12,20-27] For instance, while the Ln-doped $UO_2$ usually preserves the fluorite *Fm-3m* structure as Ln substituting U in the 4*b* sites, the local environment can be distorted[28] due to different charge balance mechanisms.[3,27,29] The defect chemistry then controls the oxidation behavior of Ln-doped $UO_2$.[10-12,15,21,27,30-32] Notably, the (Ln, U)$O_2$ studies usually fall into two categories in terms of change of variable, including $UO_2$ doped with different types of Ln and sample fabricated in the same condition,[12] or doping $UO_2$ by one type of Ln under different $P_{O2}$.[26,27,33]

Thus, recognizing both Ln type and preparation condition (e.g., oxygen partial pressure/$P_{O2}$) affect resulted defect structure, which then translate to physical and chemical properties,[10,12,26,34,35] we summarize and propose a revised defect chemistry description of (Ln,U)$O_2$ (**Figure 1**). The "End point 1" and "End point 2" are for charge balance mechanisms of incorporating trivalent Ln but only forming oxygen vacancy ($V_{\ddot{O}}$) and $U^{5+}$, respectively. With End point 2 being always the stoichiometric, Region 1 and Region 2 are for hypostoichiometry (O vs (U+Ln)<2), and Region 3 (right to End point 2) are for hyperstoichiometry (O vs. (U+Ln)>2). Particularly in Region 2, $U^{5+}$ and $O_i''$ co-exist to jointly compensate the charge balance. Then in Region 3, $U^{5+}$ and $O_i''$ compensates each other under the high oxygen partial pressure conditions. (Ce, U)$O_2$ shows a similar charge compensation mechanism with an extra variable of $Ce^{3+}$.

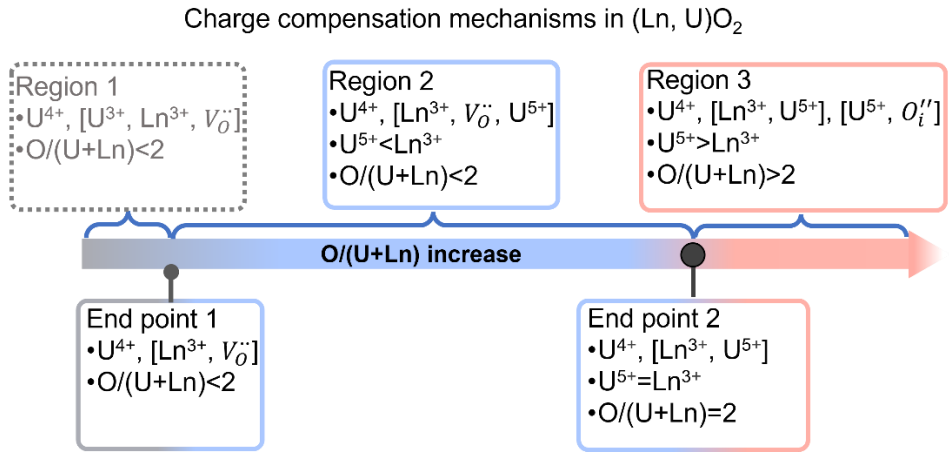

**Figure 1.** Defect chemistry diagram of (Ln, U)$O_2$ involving three Regions and two Endpoints.

This revised defect diagram of (Ln ,U)$O_2$ allows us to phenomenologically describe and predict how the charge balance mechanism and later the oxidation behavior depends on type of Ln and $P_{O2}$ during fabrication. Two questions around the resulted (Ln, U)$O_2$ defect structures are answered in this study: (1) what are the local structures of U and guest Ln in (Ln, U)$O_2$ in the non-stoichiometric phase, and (2) how the formed local structures impact the thermal oxidation pathways. Knowing these enables us to predict the oxidation behavior of SNF and enhanced fuel or MOX under various storage or off-normal conditions. Moreover, this work also has an implication for nuclear safeguard that if the fabrication condition of particular (Ln, U)$O_2$ has a distinct associated defect structure and thus correspond to a particular thermal degradation pattern, it can be used as fingerprints of $UO_2$ fuel material for tracking the originate of the fuel, burnup values or MOX stages, and fabrication history, serving for nuclear forensic purposes.[36,37]

**Results**

Three Ln-type samples were prepared: $(U_{0.51}Ce_{0.49})O_2$, $(U_{0.80}Nd_{0.20})O_2$, and $(U_{0.75}Gd_{0.25})O_2$ (here the oxygen stoichiometry is not the actual one; just for labeling sample), all of which have been calcinated in reduced (Ar and $H_2$) or nonreduced (Ar only) conditions (**Abstract Figure**). U and Ln $L_{III}$-edge X-ray absorption fine structure (XAFS) have been conducted and used for identifying the local environment of both U and Ln. X-ray diffraction (XRD) determined the long-range structures. Thermogravimetric analysis (TGA) was used to study the thermal oxidation behaviors. Raman spectroscopy further confirmed the local environment distractions (e.g., loss of symmetry and O vacancies) in pristine samples and the post-TG oxidation products. The overall results (**Abstract Figure**) suggest that Ce and N forms solid solutions with U, while in $(Gd, U)O_2$, two subnano-domains were identified, with a U-Gd solid solution domain exhibiting a lower symmetry as $Ia$-3, and another Gd-poor domain remaining in the $Fm$-3$m$ symmetry.

**Figure 2** shows nine XAFS fitting results (see supplementary **Section 1** for detailed results) that reveal the local environments of U-Ln and metal-O scattering paths. Specifically, we studied the cation distribution, e.g., clustering or solid solution (used for random mixing), by using Ln vs. U CN ratio as solid solution range. Detailed in the **experimental section**, while in ideal solid solution, the Ln vs. U CN ratio equals to their molar ratio, we applied ±10% as the accuracy limitation of CN from XAFS fitting resulting in the solid solution range.[38] As shown in **Table S3** and **Figure 2a**, the Ce vs. U ratios from the in U (4.5/4.9=0.92 for reduced and 5.2/6.6=0.79 for nonreduced) and Ce (5.0/4.5=1.11 for reduced) XAFS fall in the solid solution range (0.78 to 1.17). From either reduced or nonreduced environment, Ce and U form solid solution in the cationic sublattice. For the U-O scattering in the 1$^{st}$ shell, the O CN from the nonreduced environment (7.8 atoms) are slightly larger than that from the reduced environment (7.5 atoms), indicating oxygen vacancies favored by the later. Overall, the coordinated U has similar oxygen numbers to that of pristine $UO_2$ (**Figure 2b**). From the Ce perspective, the Ce cation is surrounded by 7.1 O atoms under the reducing condition, which is explained by $V_{\ddot{O}}$,[21,39] coupled with the formation of $Ce^{3+}$, which was observed by Ce $L_{III}$-edge XANES (see supplementary **Section 2**), Moreover, $U^{5+}$ cannot be excluded even when calcinating in reduced environment that could also form to partially compensate $Ce^{3+}$.[21] The combination of $V_{\ddot{O}}$ and $U^{5+}$ results in the slight distortion of the lattices in the reduced $(Ce, U)O_2$.

Compared to Ce, Nd introduces much server distortions in both cationic and anionic sublattices (**Figure 2c**), conventionally thought to be attributing to the trivalent state and the relatively large cation size of Nd (1.109 Å, see **Table S18**). From the U perspective, its coordinated O in the 1$^{st}$ shell has CN smaller than 8.0 atoms (5.3 atoms for reduced sample and 6.5 atoms for nonreduced sample). The missing oxygen is due to (1) the formation of $V_{\ddot{O}}$, and (2) large Debye-Waller factors and the destructive interference between U-O sub-paths in 1$^{st}$ shell with similar distance.[40] The CN ratios of Nd vs. U are in solid solution range (0.20 to 0.30) for reduced (2.0/9.0=0.22) and nonreduced samples (1.3/6.3=0.21, see **Table S3**).[24] Notably, the U-Nd scattering in the reduced sample has a larger Debye-Waller factor ($9.4\times10^{-3}$ Å$^2$) than that of U-U scattering ($4.7\times10^{-3}$ Å$^2$). In the Nd L$_{III}$-edge XAFS fitting from the reduced Nd-doped UO$_2$, the CN ratio of

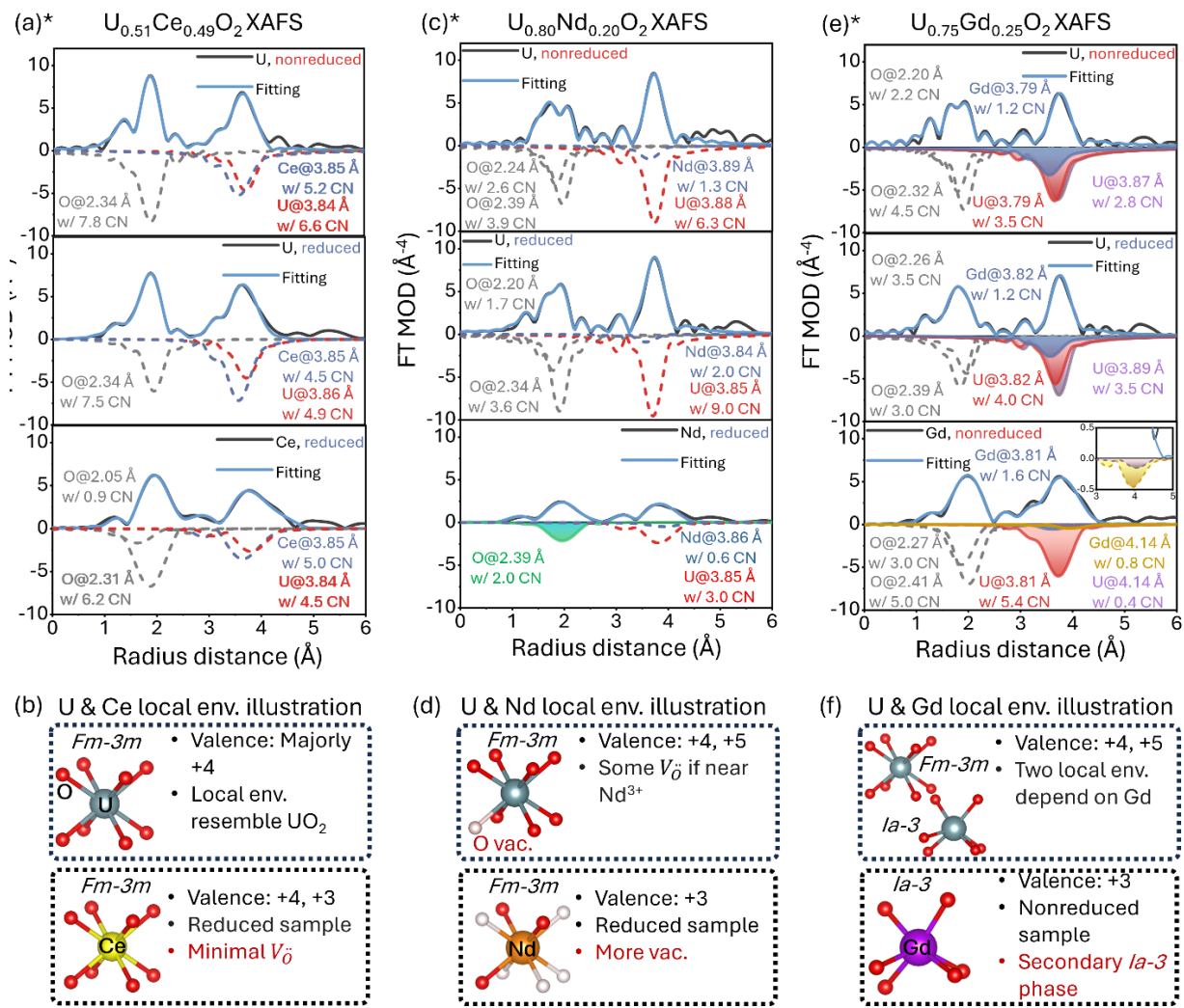

*Phase correction was not applied

**Figure 2.** Local environment analysis of Ce-, Nd- and Gd-incorporated UO$_2$ calcinated in reduced and nonreduced environments by XAFS. (a) and (b) XAFS fitting results and U, Ce local environment illustration from (U$_{0.51}$Ce$_{0.49}$)O$_2$. (c) and (d) XAFS fitting results and U, Nd local environment illustration from (U$_{0.80}$Nd$_{0.20}$)O$_2$. (e) and (f) XAFS fitting results and U, Ce local environment illustration from (U$_{0.75}$Gd$_{0.25}$)O$_2$.

Nd-U vs. Nd-Nd also fits in solid solution range (0.6/3.0=0.20). Interestingly, there are only 2.0 O atoms at 1$^{st}$ shell suggesting that aggregation of $V_O^{\cdot\cdot}$ near Nd rather than U in the reduced sample (**Figure 2d**).

Moving to smaller Ln, Gd further disturbs the fluoride structure by inducing two types of local domains with different symmetries, namely the *Ia-3* phase with U and Gd mixing in the cationic sublattice and a Gd-poor U-only region inheriting the *Fm-3m* symmetry (see **Figures 2e** and **3f**). This can be clearly seen from U and Gd XAFS results. Two metal-metal scatterings have been identified from U centers. The shortened U-U or U-Gd scattering path is located at 3.82 Å and 3.79 Å for reduced and nonreduced samples, respectively. While the CN ratio of these two atomic pairs falls into solid solution range (see **Table 3**), they are close to the Gd-Gd distance in bixbyite $Gd_2O_3$ at 3.62 Å.[41,42] At this distance, U was also identified in the Gd $L_{III}$-edge XAFS at 3.81 Å, confirming a solid solution of U-Gd in this sublattice. Furthermore, Gd-Gd scattering path was identified at 4.14 Å, similar to that in $Gd_2O_3$, thus assuring the chosen *Ia-3* symmetry of this solid solution region. On the other hand, a longer U-metal scattering path were identified at 3.89 Å (reduced) or 3.87 Å (nonreduced) in the U $L_{III}$-edge XAFS, which is much farther than the first U-metal path, and has no corresponding Gd-U path in the Gd $L_{III}$-edge XAFS, suggesting the discovered atomic pair is exclusively U-U. Overall, these results indicate two local environments in $(Gd, U)O_2$, with a *Ia-3* domain mimicking $Gd_2O_3$ and a *Fm-3m* domain just like pristine $UO_2$ (**Figure 2f**).

Although two local domains were identified in $(Gd, U)O_2$ samples, the global symmetry remains as *Fm-3m*. $(Ce, U)O_2$ and $(Nd, U)O_2$ also crystallized in the fluorite structure. These conclusions were confirmed by high resolution XRD (**Figures S11** to **S13** and **Table S4**). Furthermore, high temperature (HT from room temperature to 807 °C) XRD were performed on reduced $(Ce, U)O_2$ and reduced/nonreduced $(Nd, U)O_2$ to examine the impact of calcination condition on local environment (see supplementary **Section 3**). The temperature-dependent atomic displacement parameters of reduced $(Nd, U)O_2$ were found to be the highest, even at RT (**Figure S6**), suggesting the most static disordering with introducing Nd under the reducing condition.

Oxygen vacancies and defects can be better examined by Raman spectroscopy, accessible by two Raman bands (**Figure 3a**), normalized $V_O$ (about 540 cm$^{-1}$, related to $V_{\ddot{O}}$) and 1LO band (560 to 570 cm$^{-1}$ relate to broken symmetry).[43] The reduced $(Nd, U)O_2$ exhibits a larger area of $V_O$ (1.00) and 1LO band (2.73) than that of nonreduced sample (0.90 and 1.88). This is consistent with XRD analysis that Nd under the reduction condition introduces a higher number of $V_{\ddot{O}}$ and further impact Nd static disordering. The reduced $(Ce, U)O_2$ also contains a higher number of $V_{\ddot{O}}$ than that of nonreduced sample due to the larger $V_O$ band (1.02 vs. 0.80), but both reduced and nonreduced samples share similar degree of distortion due to similar 1LO band areas (1.50 vs. 1.57). Then the reduced $(Gd, U)O_2$ exhibits a smaller 1LO band and $V_O$ band (0.67 and 1.14) than those in the nonreduced (0.93 and 1.80) suggesting that the nonreduced sample exhibits an averaged lower symmetry. It is possible that the portion of *Ia-3* phase in the nonreduced (Gd, U)$O_2$ is higher than the reduced one.

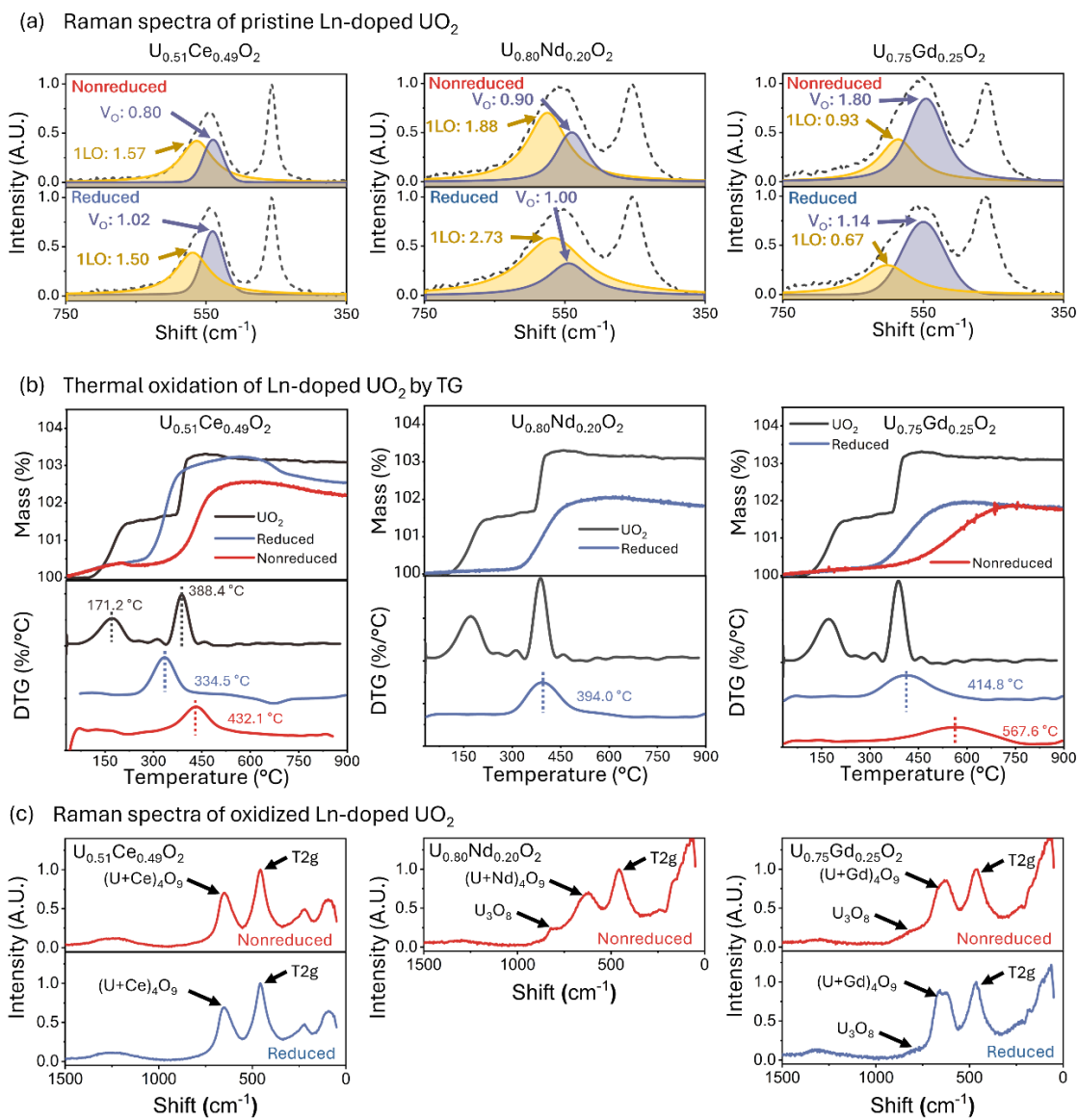

**Figure 3.** Thermal oxidation behavior of (Ln, U)$O_2$ observed by TG and change of local environment features studied by Raman spectroscopy. (a) Differences of local environment features, 1LO band and $V_O$ band area vs. $T_{2g}$ band, in Ln-doped $UO_2$ before thermal oxidation. See Table S20 for peak area details. (b) Thermal oxidation of Ln-doped $UO_2$ in synthetic air up to 900 °C observed by TG. (c) Identification of oxidation products, including (U+Ln)$_4$O$_9$ and $U_3O_8$, by Raman spectroscopy.

Lastly, the impact of the local environments on thermal oxidation was observed by thermogravimetric analysis. TG and differential thermogravimetric (DTG) analyses clearly show the enhancement of thermal oxidation due to the introduction of Ln (**Figure 3b**). The pristine $UO_2$ exhibits a two-stage oxidation pattern.[44] The first oxidation peaks at 171.2 °C, corresponding to the oxidation of $UO_2$ to $U_4O_9$. The second stage, peaking at 388.4 °C, corresponds to the oxidation of $U_4O_9$ to $U_3O_8$ through immediate $U_3O_7$. In contrast to the two-stage thermal oxidation, all (Ln, U)$O_2$ exhibit only one major oxidation peak, consistent to past studies.[12,14,15] Additionally, previous experimental studies also show that the oxidation temperatures of the doped $UO_2$ are generally influenced by the type of Ln and ratio of dopant vs. U.[12,45-47]

The oxidation temperatures of reduced (Ln, U)$O_2$ are close to the second oxidation temperature step of pristine $UO_2$. For instance, the reduced $(U_{0.51}Ce_{0.49})O_2$, $(U_{0.80}Nd_{0.20})O_2$, and $(U_{0.75}Gd_{0.25})O_2$ exhibit the peak oxidation temperature at 334.5 °C, 395.1 °C, and 414.8 °C, respectively. The nonreduced samples overall have higher oxidation temperatures than their reduced counterparts. Nonreduced $(U_{0.51}Ce_{0.49})O_2$ and $(U_{0.75}Gd_{0.25})O_2$ oxidized at 432.1 °C and 567.6 °C, respectively. Because of possible both $V_{\ddot{O}}$ and $U^{5+}$ charge balancing the guest trivalent Ln (in other word in Region 2 in **Figure 1**), it is challenging to determine the exact final oxidation products solely by TG analysis. Instead, we obtained a potential O vs. U and O vs. (U+M) range for (Ln, U)$O_2$ (see discussion section), and use that as a platform for evaluating thermal oxidation products.

Raman spectroscopy was further used to help determine the oxidation products (**Figure 3c**). The strong peak at about 650 cm$^{-1}$, signaturing $U_4O_9$,[48] suggests both $(U_{0.51}Ce_{0.49})O_2$ oxidizing to (Ce, U)$_4O_9$. Previously, the $U_4O_9$ phase was also identified in the $UO_2$-$CeO_2$ system during oxidation.[32] The oxidized (Nd, U)$O_2$ and (Gd, U)$O_2$ samples contain a mixture of (Ln, U)$_3O_8$ and (Ln, U)$_4O_9$ phases due to the presence of peaks at about 650 cm$^{-1}$ as well as above 800 cm$^{-1}$, the distinctive feature of $U_3O_8$.[49] Since trivalent Ln have low solubility in $U_3O_8$,[50] it is expected that high Ln content prohibits the formation of (Ln, U)$_3O_8$. The maximum overall cation valences after the thermal oxidation (Nd-doped: 4.8, Gd-doped: 4.7) are below +5.3 in $U_3O_8$, see **Table S22**). In the discussion, we will explore the possible connection of defect structures of starting (Ln, U)$O_2$ with the final oxidation products (e.g., $U_4O_9$, $U_3O_8$ and others).

**Discussion**

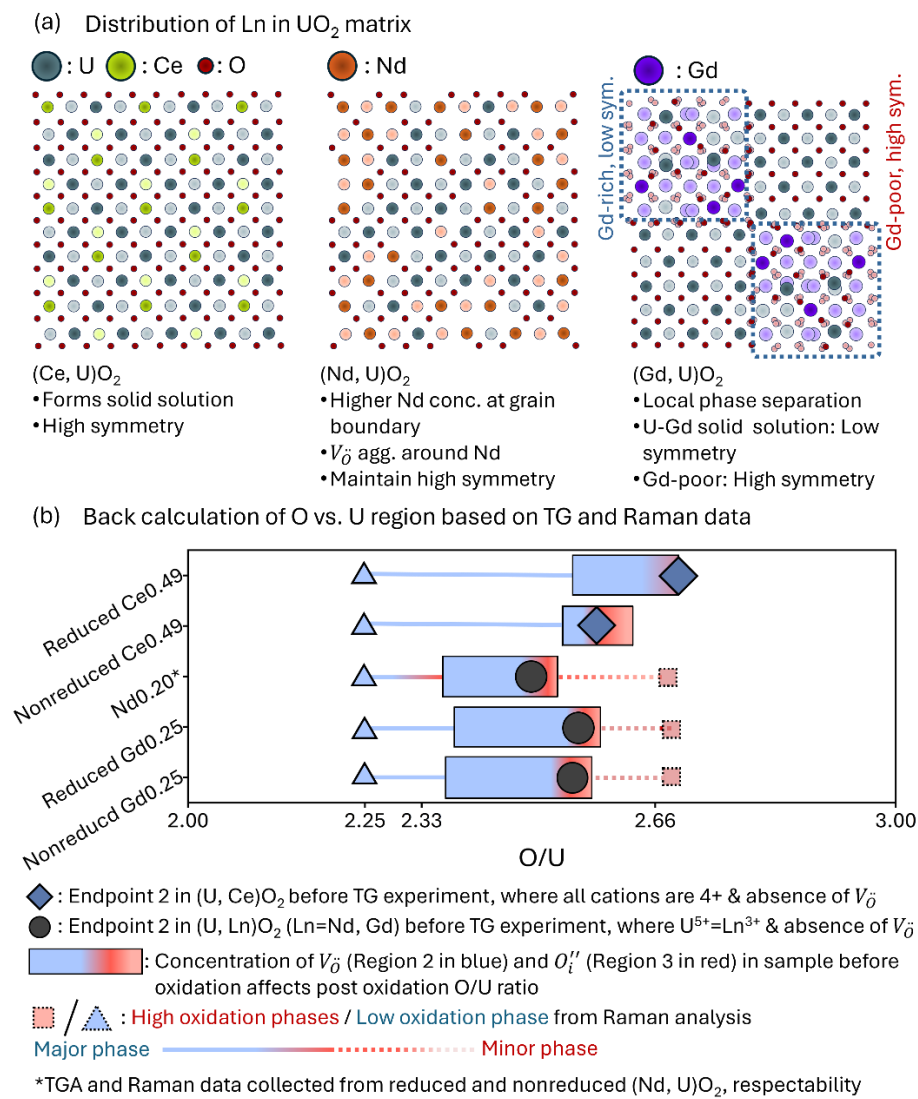

**Figure 4.** Charge compensation mechanism and Ln distribution in UO$_2$ matrix in pristine (Ln, U)O$_2$. (a) Illustration of Ln distribution in UO$_2$ matrix. The short-range low symmetry phase in (Gd, U)O$_2$ is highlighted by blue box. (b) Back calculation of charge compensation mechanism of pristine (Ln, U)O$_2$ in terms of O vs. U ratio where Ln oxide portion is excluded. For reduced (Ce, U)O$_2$, the left side of bar presents 30% of Ce cations are in +3 status, and $V_{\ddot{O}}$ is the dominate charge compensation mechanism. The right side of the bar indicates all U and Ce cations are in +4 valence. For nonreduced (Ce, U)O$_2$, the left side of the bar presents 10% of Ce cations are in +3 status, while all U cations are in +4 status. The right side of the bar shows 10% of U cations are in +5 status, while all Ce cations are in +4 status. For Nd and Gd, the left side of the bars represents Endpoint 1 in pristine sample, while the right side of the bars represents an extra 10% of U are oxidized to +5 in the pristine sample beyond Endpoint 2. See **Table S22** to **S24**, **Figure S15**, and supplementary **Section 5** for detailed calculation procedures.

Our results show that type of Ln and preparation condition of (Ln, U)O$_2$ both influence defect chemistry and then thermal oxidation behavior. Below we will explain such correlations from three perspectives, including local environment (first CN and short-range domain segregation), existence of multiple oxidation products, and possible kinetic models or thermodynamic rationales.

**Local environment of (Ln, U)O$_2$.** (Ce, U)O$_2$, regardless of the calcination conditions, exhibits a solid solution between U and Ce in the cationic sublattice, based on the U and Ce L$_{III}$-edge XAFS analyses (**Figure 4a**). This is mostly due to the major valence state to be tetravalent instead of trivalent for Ce in UO$_2$, which are compatible with U$^{4+}$ in terms of size and charge. Although a small amount of Ce$^{3+}$ was observed (see **supplementary Section 2**), the equilibrium sample will likely to be dominated by Ce$^{4+}$-U$^{4+}$ theme to be the most thermodynamically favorable, which also suggests a close-to-ideal mixing for U and Ce.[51] The overall degree of oxygen defects (e.g., vacancy and O clusters) between the two samples are similar as they share 1LO bands with similar area in Raman (**Table S20**), .[52] The reduced sample does have a higher concentration of $V_{\ddot{O}}$, confirmed by Raman and later by thermal oxidation studies. For (Nd, U)O$_2$, although Nd were confirmed to occupy the crystallographic 4$b$ site within the *Fm-3m* structure, the induced $V_{\ddot{O}}$ (from the reduced condition) was found to be mostly aggregated around Nd instead of U, leading to a large static disordering of the center Nd at the 4$b$ site (see supplementary **Section 4** for details), due to the releases of geometric constrain. This finding from HT XRD is consistent with the Raman data where the reduced sample has a larger 1LO distortion band than that of nonreduced sample. In contrast, the defect mechanism shifts when moved to the nonreduced sample that have less $V_{\ddot{O}}$ with a more ordered Nd at the 4$b$ site. In other words, one will find reduced (Nd, U)O$_2$ in Region 2 toward End point 1 while the nonreduced also in Region 2 but toward End point 2 in the defect chemistry diagram (**Figure 1**).

When late lanthanide with smaller size incorporated into UO$_2$, and here we used Gd as the representative element for late trivalent Ln, we begin to see deviations in Ln-dominated site vs U sites in the global fluorite structure, namely the original *Fm-3m* (higher symmetry) domain and the *Ia-3* (lower symmetry seen in the bixbyite structure) domain. A cationic solid solution of U-Gd was observed in the *Ia-3* domain, while the *Fm-3m* occupying almost exclusively U cation. We hypothesized the *Ia-3* domain to be stoichiometry in composition so the defect chemistry can be described by End point 2: (Gd$^{3+}$, U$^{5+}$)O$_2$ with a *Ia-3* atomic configuration. Both reduced and nonreduced samples shows such "domain separation" and the differences between them are small: the nonreduced sample contains a higher ratio of low symmetry sites due to the high concentration of U$^{5+}$ that charge coupled with Gd$^{3+}$ forming the low symmetry domain. The reason of forming the *Ia-3* domain in (Gd, U) O$_2$, instead of (Nd, U) O$_2$, may relate to the thermodynamic stabilities of their sesquioxide phases. Nd$_2$O$_3$ has the *hexagonal P-3m*1 phase as the most stable phase under the standard condition, while the Gd$_2$O$_3$ stabilizes in the *cubic Ia*-3 phase.[41,53] Moreover, the enthalpy of formation per mole of atom of Gd$_2$O$_3$ (-363.9 kJ/mol) is more negative than that of UO$_2$ (-361.7 kJ/mol), while Nd$_2$O$_3$ exhibits the highest formation enthalpy (-361.4 kJ/mol) which also has crystallographic mismatch with that of $UO_2$.[54] Therefore, the formation of $(Gd^{3+}, U^{5+})O_2$ *cubic Ia-3* domain is thermodynamically favorable and crystallographically compatible with the hosting $UO_2$ matrix. From this thermodynamic perspective, such *Ia*-3 domain is also expected to exist in late Ln-incorporated $UO_2$ because all late $Ln_2O_3$ have stable phases crystallized in cubic *Ia*-3 and their enthalpies of formation are more favorable than that of $UO_2$.[42,55] For instance, the stable phase of $Yb_2O_3$ (Yb being the very late lanthanide) is also *Ia*-3, with an favorable enthalpy of formation per mole of atom (-362.9 kJ/mol) than that of $UO_2$,[54] making (Yb, U)$O_2$ also prone to form *Ia*-3 domains.

Noted here even we refer the two domains with two distinct crystallographic motifs, the low symmetry *Ia*-3 does not extend to the long range to be XRD visible, where from XRD we only observed and confirmed one structure and that is *Fm*-3*m* fluorite structure. Therefore, this is one case of "long-range disordering, short-range ordering", where the local symmetries do not translate completely to the global one. This is not uncommon and have been found previously but in a limited studies, including on zircon[56], pyrochlore[57,58], $ZrO_2$.[59] In **Figure 4a**, we schematically show the two symmetry phases pack among themselves with one phase taking one unit-cell as domain boundary that is also about the highest limit of what XAFS can probe. Clearly seen from the two dimensional 2 × 2 supercell, (Ce, U)$O_2$ and (Nd, U)$O_2$ both have cationic solid solutions, while Nd incorporation induces considerable amounts of $V_{\ddot{O}}$. (Gd, U)$O_2$ have the *Fm*-3*m* domains and the *Ia*-3 domains that align along the domain walls. The diminish of the *Ia*-3 symmetry can occur when the associate domains configurationally arrange so disordering of lower symmetry generates a high symmetry pattern.

**Thermal oxidation of (Ln, U)$O_2$.** With only the defect chemistry qualitatively described in **Figure 1**, we can only determine the composition range of the final thermal oxidation products. This can be done by setting a range of defect regions in the pristine samples and using them as starting compositions. The calculated range of final compositions are plotted in **Figure 4b** for O vs. U (after excluding O from Ln oxides) and **Figure S16** for O vs. (U+Ln). The bars show how the concentration of $V_{\ddot{O}}$ in Region 2 (blue) and $O_i''$ in Region 3 (red) in pristine samples impact the post-oxidation O vs. U ratio based on current TG data, while the lines, cubes, and triangles indicate the phases detected from Raman. Based on our TGA data, the estimated final composition from (Ce, U)$O_2$ at Endpoint 2 resulted the final U phases to have a O/U = 2.66 corresponding to mass equivalent of $U_3O_8$ (**Table S22**). This result shows that Ce is not effective in terms of oxidation prevention than Nd and Gd (discussed later) potentially due to its tetravalent nature. The above TG analysis assumes pure phase composition in the product, which could be the averaged composition if multiple phases are present. From Raman, both (Ce, U)$O_2$ samples only oxidized to (Ce, U)$_4O_9$. This result in a fair agreement with TGA when translating the O/U=2.66 to O/(U+Ln) with consideration of presence of trivalent Ce (estimated to be 10 ~ 30 % in the left side of the bar in blue region,

**Figure 4, Table S24**) resulted in 2.27 to 2.29 that is close to the (Ce, U)$_4$O$_9$ oxygen:metal ratio. Considering the high Ce ratio (0.49) and random mixture nature between Ce and U, Ce significantly affects the oxidation of U by preventing the aggregation and rearrangement of $O_i''$ clusters and subsequent cation sublattice rearrangement,[60-63] thus affect the formation of final oxidation products.

Trivalent Nd and Gd are more effective in preventing oxidation (**Figure 3**). TGA suggests the O vs. U ratios in the oxidation products if samples start at Region 2 ($V_{\ddot{O}}$ and U$^{5+}$ exist simultaneously, the most likely case) are between 2.33 and 2.50. This O/U ratio suggests the final products could be a mixture of U$_4$O$_9$ or U$_3$O$_7$ with U$_3$O$_8$, which is consistent with Raman results where we observed a large portion of (Ln, U)$_4$O$_9$ and small portion of (Ln, U)$_3$O$_8$.

**Proposed kinetic models with short-range features to explain oxidation resistance of (Ln, U)O$_2$.** The difference of thermal oxidation pathways between UO$_2$ and (Ln, U)O$_2$ have been previously explained by Nawada et al., Kim et al., Ha et al., De Alleluia et al., and McEachern et al. in terms of size and charge due to the less accessibility of Ln to higher valences,[64-68] and by Vinograd et al. with thermodynamic models.[10,11] The existing models can rationalize the thermal oxidation of (Ce, U)O$_2$, where Ce and U forms *Fm-3m* solid solutions. The enhanced thermal oxidation resistance is due to the inability of Ce oxidation beyond 4+ and prevents the formation of $O_i''$ nearby.[60-63] The higher oxidation temperature seen in nonreduced (Ce, U)O$_2$ is due to the lower concentrations of $V_{\ddot{O}}$. With some of the past models capable of explaining the delays in oxidation, they are unable to rationalize the trend seen from early to late trivalent lanthanides. Here we propose a new and simplified kinetic model based on the local defect structures (see **Figure 4a**) from this work. The kinetic model involves two-step reactions: (1) filling $V_{\ddot{O}}$ near Ln$^{3+}$ to form (Ln$^{3+}$, U$^{5+}$)O$_2$ clusters (and to *Ia*-3 domains if reaction extends), with low activation energies, and followed by (2) insertion of $O_i''$ into the *Fm-3m* structure for further oxidation, with higher activation energies.

This model explains the enhanced thermal oxidation by applying it to local structures and TGA results of (Nd, U)O$_2$ and (Gd, U)O$_2$, which represent early and late Ln incorporation, respectively. When (Nd, U)O$_2$ starts oxidizing, the $V_{\ddot{O}}$ near the Nd centers are filled firstly with nearby U$^{4+}$ also oxidized to U$^{5+}$, which requires only low activation energies provided from heating, thus this can occur in lower temperatures (394 °C). The sequential oxidation with insertion of $O_i''$ plausibly occurs coincidentally. (Nd$^{3+}$, U$^{5+}$)O$_2$ *Ia*-3 domains may also form after the first-step oxidation, which assumes to carry higher activation energies than the UO$_2$ domains during the second-step $O_i''$-insertion oxidation. In this case, again due to the inability of Nd$^{3+}$ to be further oxidized to higher valence state and high valence of U$^{5+}$, the low-symmetry domain was hypothesized to be oxidized to phase close to U$_4$O$_9$. The UO$_2$ domains, however, can be oxidized to U$_3$O$_8$. It is in fact both phases were observed in Raman, with their corresponding bands identified.

For (Gd, U)O$_2$, the pre-existence of the short-range *Ia*-3 (rather than the product from the step 1) is the key for improving the thermal oxidation resistance. Without too many $V_{\ddot{O}}$ charge-coupling Gd$^{3+}$, the (Gd$^{3+}$, U$^{5+}$)O$_2$ has little oxygen vacancies to be filled up, thus attenuating the first-step oxidation. As the second step seeks higher thermal energy to activate, (Gd, U)O$_2$ oxidized in high temperatures at 415 ~ 568 °C. The slightly lower temperature (415 °C) for the reduced (Gd, U)O$_2$ also suggests the minor first-step oxidation where filling $V_{\ddot{O}}$ still occurs but encounters slightly higher energetic barriers (compared to 394 °C for Nd case). In the nonreduced sample where $V_{\ddot{O}}$ is minimized, the second-step oxidation dominates. Existing of the two domains stabilizes the integrity of (Gd, U)O$_2$ close to 500 °C, then with the *Ia*-3 domains continuingly resistant to thermal oxidation and the *Fm*-3*m* domains more capable of being oxidized to higher-valence uranium oxide phases, similar to the second-stage oxidation of (Nd, U)O$_2$. Again, we observed both (Gd, U)$_4$O$_9$ and (Gd, U)$_3$O$_8$ phases from the related Raman signatures. Although we cannot pinpoint the exact compositions for each oxidation phases, future work by micro-XRD and micro-XRF would be appropriate to continue determining the solubility of Ln in these higher-valence U oxide phases.

This kinetic model can further applied to our previous studiy on (Yb, U)O$_2$ that showed excellent thermal oxidation resistance (peaked at ~700 °C).[12] Based on previous discussion, (Yb, U)O$_2$ also favors the formation of *Ia*-3 domains and limits the formation of $V_{\ddot{O}}$. Hypothetically, such similar short-range motifs also contribute to the better thermal oxidation resistance of (Yb, U)O$_2$. Overall, this kinetic model, despite being over-simplified, connects well the phenomenological defect chemistry schematics with thermal oxidation behaviors, which particularly in combination explains why late lanthanide incorporation can generate benefits for thermal oxidation resistance.

**Hypothetical thermodynamic explanation for Ln-U oxide systems and their roles in thermal oxidation.** However, as what Vinograd proposed in his thermodynamic modeling work on the (Ln, U)O$_2$ solid solutions,[10,11] we believe beneath the kinetic model, what dedicates the oxidation behaviors is thermodynamics, and specifically, the thermodynamic parameters of solid solution phases of Ln with U in UO$_2$, and all related thermal oxidation phases (e.g., UO$_{2+x}$, U$_4$O$_9$, U$_3$O$_7$, and U$_3$O$_8$). Knowing the enthalpies of formation and mixing of each Ln-U solid solutions will generate thermodynamic pictures to evaluate solubility and stability of Ln in the hosting U oxide phases. It has been know that non-ideal mixing will lead to enthalpically stabilization of phases that are unstable in endmember composition.[56,69] U$_4$O$_9$ or U$_3$O$_7$, traditionally knows as metastable phases, could possibly be stabilized by including Ln, which can lead to phase competition with Ln-incorporated U$_3$O$_8$ that seems to be destabilized due to the incompatibility of Ln$^{3+}$ in high-valence sites. This may explain why the U$_4$O$_9$-like phase formed as the major phase in the final oxidation mixture. To make such an assessment from hypothetical to factual, we need to know all the related thermodynamics parameters, whose availability, unfortunately, are extremely limited. Vinograd has tried to model $Ln_2O_3$-$UO_2$-$UO_{2.5}$ systems with Margules interaction parameters empirically obtained or hypothetically obtained, which could have underestimated the repulsion interaction between $Ln^{3+}$ with $U^{4+}$ (and $U^{5+}$) in the fluorite structure. Although Zhang et al. have calorimetrically measured Ln-doped (Ln = La, Y, Nd) $UO_2$ systems,[13] due to the ambiguity of the defect chemistry of the studied samples, no interaction parameters (assuming regular solution model) were obtained. The interaction parameters existing in $UO_2$ with other tetravalent oxide systems such as $CeO_2$,[51] and actinide oxides,[70] obtained mostly computationally but cannot translate to $Ln_2O_3$-$UO_2$ systems due to major differences in defect chemistry. Thus, we conclude that up to date, we have almost no benchmarked interaction parameters to properly model the $Ln_2O_3$-$UO_2$ systems. Even larger knowledge gaps exist in understanding mixing Ln with U in higher-valence U oxide phases. As a perspective for future work, we call out such pressing research needs in nuclear chemistry community, and we will be committed to perform systematic calorimetric studies on these solid solution systems.

**Conclusions and Implications**

This study proposed a new defect chemistry diagram for describing (Ln, U)$O_2$ systems, coupled with a simplified kinetic model of thermal oxidation. The formed (Ln, U)$O_2$ are non-stoichiometric with the defect structures (e.g., concentration of oxygen vacancies, $U^{5+}$, and short-range features) controlled by Ln type and preparation conditions. From early to late lanthanides, new low-symmetry *Ia*-3 domain has been discovered and assumed to be dominated with introducing smaller Ln. The *Ia*-3 domain is subnano-sized and, with configurational arrangement, is unable to translate to global symmetry that remains as *Fm*-3*m*. Existing of the subnano-domain plays a critical role in improving thermal oxidation resistance due to the inhabitation of low-energy activation oxidation step. We also highlights the importance of thermodynamic properties of various Ln-U oxide systems (from (Ln, U)$O_2$ to (Ln, U)$_3O_8$. This is crucial for better understanding the formation of Ln-rich secondary phase in the $UO_2$ matrix and predicting the oxidation resistance of (Ln, U)$O_2$.

From the practical application perspective, these findings contribute significantly to two key areas: nuclear forensics and fuel design. By demonstrating the sensitivity of the local environment in (Ln, U)$O_2$ to fabrication conditions, this study offers a potential fingerprint for identifying the originate of nuclear materials, aiding in nuclear non-proliferation efforts.[37] The non-equilibrated nature of formed (Ln, U)$O_2$ particles, which is strongly influenced by fabrication environment and fission product inclusion (other word, burnup values), can provide "fingerprint" in their local structure, accessible by XAFS and Raman, or thermal fingerprint when examined by thermal oxidation analysis. Predictable tools such as improved and quantitative defect chemistry diagram with analytical TGA scales based on those developed in this work will help regulators use for identifying origin of sample fabrication, and tracing sample transfer history. On the front-end of nuclear industry, the enhanced oxidation resistance confirmed and rationalized in (Ln, U)O$_2$ suggests potential benefits for the development of more stable and accident-tolerant nuclear fuels. It also aligns with the potential benefits of incorporating Ln-based burnable absorbers into UO$_2$ fuel.[71] Within a reactor, burnable absorbers capture excess neutrons, moderating the fission reaction and extending fuel life. Medium and late Ln elements are particularly attractive candidates due to their high neutron absorption cross-sections. While our study demonstrates the additional benefits of including these Ln that is improved thermal oxidation resistance.

**Acknowledgements**

This work was supported by the National Science Foundation (NSF), Division of Materials Research, under award No. 2144792, and the United States Nuclear Regulatory Commission, Office of Nuclear Regulatory Research (RES) under award No. 31310023M0011. Portions of this research were supported by collaboration, services, and infrastructure through the Nuclear Science Center User Facility at WSU, the WSU-PNNL Nuclear Science and Technology Institute, and Alexandra Navrotsky Institute for Experimental Thermodynamics. This work was performed at 20-BM, Advanced Photon Source, Argonne National Laboratory. This research used resources of the Advanced Photon Source, a U.S. Department of Energy (DOE) Office of Science user facility operated for the DOE Office of Science by Argonne National Laboratory under Contract No. DE-AC02-06CH11357. This research used resources 6-BM and 28-ID-B of the National Synchrotron Light Source II, a U.S. Department of Energy (DOE) Office of Science User Facility operated for the DOE Office of Science by Brookhaven National Laboratory under Contract No. DE-SC0012704.

**Disclaimer**

Prepared by Xiaofeng Guo under award No. 31310023M0011 from Office of Nuclear Regulatory Research (RES), Nuclear Regulatory Commission, the statements, findings, conclusions, and recommendations in this paper are those of the author(s) and do not necessarily reflect the view of the RES or the US Nuclear Regulatory Commission

# Reference

1. Zinkle, S. J. & Was, G. S. Materials challenges in nuclear energy. *Acta Materialia* **61**, 735-758 (2013). https://doi.org:https://doi.org/10.1016/j.actamat.2012.11.004
2. Yun, D. *et al.* Current state and prospect on the development of advanced nuclear fuel system materials: A review. *Materials Reports: Energy* **1**, 100007 (2021). https://doi.org:https://doi.org/10.1016/j.matre.2021.01.002
3. Bès, R. *et al.* New insight in the uranium valence state determination in UyNd1− yO2±x. *Journal of Nuclear Materials* **507**, 145-150 (2018).
4. Shadrin, A. Y. *et al.* Fuel fabrication and reprocessing for nuclear fuel cycle with inherent safety demands. *Radiochimica Acta* **103**, 163-173 (2015). https://doi.org:doi:10.1515/ract-2015-2385
5. Cooper, M. W. D., Stanek, C. R. & Andersson, D. A. The role of dopant charge state on defect chemistry and grain growth of doped UO2. *Acta Materialia* **150**, 403-413 (2018). https://doi.org:https://doi.org/10.1016/j.actamat.2018.02.020
6. Smith, H., Cordara, T., Gausse, C., Pepper, S. E. & Corkhill, C. L. Oxidative dissolution of Cr-doped UO2 nuclear fuel. *npj Materials Degradation* **7**, 25 (2023). https://doi.org:10.1038/s41529-023-00347-4
7. Bourgeois, L., Dehaudt, P., Lemaignan, C. & Hammou, A. Factors governing microstructure development of Cr2O3-doped UO2 during sintering. *Journal of Nuclear Materials* **297**, 313-326 (2001). https://doi.org:https://doi.org/10.1016/S0022-3115(01)00626-2
8. Cooper, M. W. D. *et al.* Fission gas diffusion and release for Cr2O3-doped UO2: From the atomic to the engineering scale. *Journal of Nuclear Materials* **545**, 152590 (2021). https://doi.org:https://doi.org/10.1016/j.jnucmat.2020.152590
9. Cardinaels, T. *et al.* Chromia doped UO2 fuel: Investigation of the lattice parameter. *Journal of Nuclear Materials* **424**, 252-260 (2012). https://doi.org:https://doi.org/10.1016/j.jnucmat.2012.02.025
10. Vinograd, V. L., Bukaemskiy, A. A., Modolo, G., Deissmann, G. & Bosbach, D. Thermodynamic and Structural Modelling of Non-Stoichiometric Ln-Doped UO(2) Solid Solutions, Ln = La, Pr, Nd, Gd. *Front Chem* **9**, 705024 (2021). https://doi.org:10.3389/fchem.2021.705024
11. Vinograd, V. L., Bukaemskiy, A. A., Deissmann, G. & Modolo, G. Thermodynamic model of the oxidation of Ln-doped UO(2). *Sci Rep* **13**, 17944 (2023). https://doi.org:10.1038/s41598-023-42616-x
12. Olds, T. A., Karcher, S. E., Kriegsman, K. W., Guo, X. & McCloy, J. S. Oxidation and anion lattice defect signatures of hypostoichiometric lanthanide-doped UO2. *Journal of Nuclear Materials* **530**, 151959 (2020). https://doi.org:https://doi.org/10.1016/j.jnucmat.2019.151959
13. Zhang, L. & Navrotsky, A. Thermochemistry of rare earth doped uranium oxides LnxU1−xO2−0.5x+y (Ln = La, Y, Nd). *Journal of Nuclear Materials* **465**, 682-691 (2015). https://doi.org:https://doi.org/10.1016/j.jnucmat.2015.06.059
14. Karcher, S. *et al.* Benefits of using multiple Raman laser wavelengths for characterizing defects in a UO2 matrix. *Journal of Raman Spectroscopy* **53**, 988-1002 (2022). https://doi.org:https://doi.org/10.1002/jrs.6321
15. Scheele, R. D., Hanson, B. D. & Casella, A. M. Effect of added gadolinium oxide on the thermal air oxidation of uranium dioxide. *Journal of Nuclear Materials* **552**, 153008 (2021). https://doi.org:https://doi.org/10.1016/j.jnucmat.2021.153008
16. Veliscek-Carolan, J. Separation of actinides from spent nuclear fuel: A review. *Journal of Hazardous Materials* **318**, 266-281 (2016). https://doi.org:https://doi.org/10.1016/j.jhazmat.2016.07.027
17. Wang, Z. *et al.* Ultra-Efficient Americium/Lanthanide Separation through Oxidation State Control. *Journal of the American Chemical Society* **144**, 6383-6389 (2022). https://doi.org:10.1021/jacs.2c00594

18  Fouquet-Métivier, P. *et al.* Insight into the Cationic Charge Distribution in U1–y–zPuyAmzO2±x Mixed Oxides. *Inorganic Chemistry* **63**, 20482-20491 (2024). https://doi.org:10.1021/acs.inorgchem.4c03084

19  Bouloré, A., Aufore, L., Federici, E., Blanpain, P. & Blachier, R. Advanced characterization of MIMAS MOX fuel microstructure to quantify the HBS formation. *Nuclear Engineering and Design* **281**, 79-87 (2015).

20  Soldati, A. L. *et al.* Synthesis and characterization of Gd2O3 doped UO2 nanoparticles. *Journal of Nuclear Materials* **479**, 436-446 (2016). https://doi.org:https://doi.org/10.1016/j.jnucmat.2016.07.033

21  Ha, Y.-K., Lee, J., Kim, J.-G. & Kim, J.-Y. Effect of Ce doping on UO2 structure and its oxidation behavior. *Journal of Nuclear Materials* **480**, 429-435 (2016). https://doi.org:https://doi.org/10.1016/j.jnucmat.2016.08.026

22  Liu, N. *et al.* Influence of Gd Doping on the Structure and Electrochemical Behavior of UO2. *Electrochimica Acta* **247**, 496-504 (2017). https://doi.org:https://doi.org/10.1016/j.electacta.2017.07.006

23  Xiao, H. *et al.* Microstructure and thermophysical properties of Er2O3-doped UO2 ceramic pellets. *Journal of Nuclear Materials* **534**, 152109 (2020). https://doi.org:https://doi.org/10.1016/j.jnucmat.2020.152109

24  Herrero, B. *et al.* Charge compensation mechanisms in Nd-doped UO2 samples for stoichiometric and hypo-stoichiometric conditions: Lack of miscibility gap. *Journal of Nuclear Materials* **539**, 152276 (2020).

25  Schreinemachers, C. *et al.* Fabrication of Nd- and Ce-doped uranium dioxide microspheres via internal gelation. *Journal of Nuclear Materials* **535**, 152128 (2020). https://doi.org:https://doi.org/10.1016/j.jnucmat.2020.152128

26  Barral, T., Claparede, L., Podor, R. & Dacheux, N. Understanding the solid/liquid interface evolution during the dissolution of Nd-doped UO2 by macro-/microscopic dual approach. *Corrosion Science* **222**, 111380 (2023). https://doi.org:https://doi.org/10.1016/j.corsci.2023.111380

27  Barral, T. *et al.* Impact of the atmosphere on the sintering capability and chemical durability of Nd-doped UO2+x mixed oxides. *Journal of the European Ceramic Society*, 116845 (2024). https://doi.org:https://doi.org/10.1016/j.jeurceramsoc.2024.116845

28  Desgranges, L. *et al.* Understanding Local Structure versus Long‐Range Structure: The Case of UO2. *Chemistry–A European Journal* **24**, 2085-2088 (2018).

29  Prieur, D. *et al.* Aliovalent Cation Substitution in UO2: Electronic and Local Structures of U1–yLayO2±x Solid Solutions. *Inorganic Chemistry* **57**, 1535-1544 (2018). https://doi.org:10.1021/acs.inorgchem.7b02839

30  Hong, M., Chun, H., Kwon, C. & Han, B. Outstanding stability of Gd-doped UO2 against surface oxidation: First-principles study. *Applied Surface Science* **589**, 152955 (2022). https://doi.org:https://doi.org/10.1016/j.apsusc.2022.152955

31  Herrero, B. *et al.* Charge compensation mechanisms in Nd-doped UO2 samples for stoichiometric and hypo-stoichiometric conditions: Lack of miscibility gap. *Journal of Nuclear Materials* **539**, 152276 (2020). https://doi.org:https://doi.org/10.1016/j.jnucmat.2020.152276

32  Nandi, C. *et al.* Phase evolution in the UO2–CeO2 system under oxidizing and reducing conditions: X-ray diffraction and spectroscopic studies. *Journal of Physics and Chemistry of Solids* **180**, 111444 (2023). https://doi.org:https://doi.org/10.1016/j.jpcs.2023.111444

33  Massonnet, M. *et al.* Influence of Sintering Conditions on the Structure and Redox Speciation of Homogeneous (U,Ce)O2+δ Ceramics: A Synchrotron Study. *Inorganic Chemistry* **62**, 7173-7185 (2023). https://doi.org:10.1021/acs.inorgchem.2c03945

34  Griveau, J.-C. *et al.* Low-temperature heat capacity and magnetism in (U1−yLny)O2 and (U1−yAmy)O2 (y = 0.01 − 0.05) solid solutions: Effects of substitution and self-irradiation. *Journal of Applied Physics* **132** (2022). https://doi.org:10.1063/5.0112674
35  Bès, R. *et al.* New insight in the uranium valence state determination in UyNd1−yO2±x. *Journal of Nuclear Materials* **507**, 145-150 (2018). https://doi.org:https://doi.org/10.1016/j.jnucmat.2018.04.046
36  Tamasi, A. L. *et al.* Oxidation and hydration of U3O8 materials following controlled exposure to temperature and humidity. *Anal Chem* **87**, 4210-4217 (2015). https://doi.org:10.1021/ac504105t
37  Pastoor, K. J., Kemp, R. S., Jensen, M. P. & Shafer, J. C. Progress in Uranium Chemistry: Driving Advances in Front-End Nuclear Fuel Cycle Forensics. *Inorg Chem* (2021). https://doi.org:10.1021/acs.inorgchem.0c03390
38  Vaarkamp, M. Obtaining reliable structural parameters from EXAFS. *Catalysis Today* **39**, 271-279 (1998). https://doi.org:https://doi.org/10.1016/S0920-5861(97)00111-9
39  Griffiths, T. R., Hubbard, H. V. S. A. & Davies, M. J. Electron transfer reactions in non-stoichiometric ceria and urania. *Inorganica chimica acta* **225**, 305-317 (1994).
40  Conradson, S. D. *et al.* Local Structure and Charge Distribution in the UO2−U4O9 System. *Inorganic Chemistry* **43**, 6922-6935 (2004). https://doi.org:10.1021/ic049748z
41  Lonappan, D. *et al.* Cubic to hexagonal structural transformation in Gd2O3 at high pressure. *Philosophical Magazine Letters* **88**, 473-479 (2008). https://doi.org:10.1080/09500830802232534
42  Zinkevich, M. Thermodynamics of rare earth sesquioxides. *Progress in Materials Science* **52**, 597-647 (2007). https://doi.org:https://doi.org/10.1016/j.pmatsci.2006.09.002
43  Elorrieta, J. M., Bonales, L. J., Rodriguez-Villagra, N., Baonza, V. G. & Cobos, J. A detailed Raman and X-ray study of UO2+x oxides and related structure transitions. *Phys Chem Chem Phys* **18**, 28209-28216 (2016). https://doi.org:10.1039/c6cp03800j
44  McEachern, R. J. & Taylor, P. A review of the oxidation of uranium dioxide at temperatures below 400 C. *Journal of Nuclear Materials* **254**, 87-121 (1998).
45  Une, K. & Oguma, M. Oxygen potentials of (U, Gd)O2 ± x solid solutions in the temperature range 1000–1500°C. *Journal of Nuclear Materials* **115**, 84-90 (1983). https://doi.org:https://doi.org/10.1016/0022-3115(83)90345-8
46  Yoshida, K. *et al.* Oxygen potential of hypo-stoichiometric La-doped UO2. *Journal of Nuclear Materials* **418**, 22-26 (2011). https://doi.org:https://doi.org/10.1016/j.jnucmat.2011.06.045
47  Osaka, M. & Tanaka, K. Oxygen potential of hypo-stoichiometric Lu-doped UO2. *Journal of Nuclear Materials* **378**, 193-196 (2008). https://doi.org:https://doi.org/10.1016/j.jnucmat.2008.06.026
48  Desgranges, L., Baldinozzi, G., Simon, P., Guimbretière, G. & Canizares, A. Raman spectrum of U4O9: a new interpretation of damage lines in UO2. *Journal of Raman Spectroscopy* **43**, 455-458 (2012). https://doi.org:10.1002/jrs.3054
49  Miskowiec, A. *et al.* Additional complexity in the Raman spectra of U3O8. *Journal of Nuclear Materials* **527**, 151790 (2019). https://doi.org:https://doi.org/10.1016/j.jnucmat.2019.151790
50  Potts, S. K. *et al.* Structural incorporation of europium into uranium oxides. *MRS Advances*, 1-7 (2023).
51  Hanken, B. E. *et al.* Energetics of cation mixing in urania–ceria solid solutions with stoichiometric oxygen concentrations. *Physical Chemistry Chemical Physics* **14**, 5680-5685 (2012).
52  Jégou, C. *et al.* Raman spectroscopy characterization of actinide oxides (U1−yPuy)O2: Resistance to oxidation by the laser beam and examination of defects. *Journal of Nuclear Materials* **405**, 235-243 (2010). https://doi.org:https://doi.org/10.1016/j.jnucmat.2010.08.005
53  Pandey, K. K., Garg, N., Mishra, A. K. & Sharma, S. M. High pressure phase transition in Nd2O3. *Journal of Physics: Conference Series* **377**, 012006 (2012). https://doi.org:10.1088/1742-6596/377/1/012006

54	Konings, R. J. M. *et al.* The Thermodynamic Properties of the f-Elements and their Compounds. Part 2. The Lanthanide and Actinide Oxides. *Journal of Physical and Chemical Reference Data* **43** (2014). https://doi.org:10.1063/1.4825256
55	Zhang, Y. & Jung, I.-H. Critical evaluation of thermodynamic properties of rare earth sesquioxides (RE = La, Ce, Pr, Nd, Pm, Sm, Eu, Gd, Tb, Dy, Ho, Er, Tm, Yb, Lu, Sc and Y). *Calphad* **58**, 169-203 (2017). https://doi.org:https://doi.org/10.1016/j.calphad.2017.07.001
56	Marcial, J. *et al.* Thermodynamic Non-ideality and Disorder Heterogeneity in Actinide Silicate Solid Solutions. *npj Material Degradation* **5**, 1-14 (2021).
57	O'Quinn, E. C. *et al.* Predicting short-range order and correlated phenomena in disordered crystalline materials. *Science Advances* **6**, eabc2758 (2020).
58	Shamblin, J. *et al.* Probing disorder in isometric pyrochlore and related complex oxides. *Nature materials* **15**, 507-511 (2016).
59	Alexandre Solomon, E. O. Q., Juejing Liu, Igor Gussev, Xiaofeng Guo, Joerg Neuefeind, Christina Trautmann, Rodney Ewing, Gianguido Baldinozzi, Maik Lang, . Atomic-Scale Structure of $ZrO_2$: Formation of Metastable Polymorphs. *Science Advances* (In press).
60	Kato, M., Watanabe, M., Hirooka, S. & Vauchy, R. Oxygen diffusion in the fluorite-type oxides $CeO_2$, $ThO_2$, $UO_2$, $PuO_2$, and $(U, Pu)O_2$. *Frontiers in Nuclear Engineering* **1** (2023). https://doi.org:10.3389/fnuen.2022.1081473
61	Leinders, G. *et al.* Assessment of the $U_3O_7$ Crystal Structure by X-ray and Electron Diffraction. *Inorganic Chemistry* **55**, 9923-9936 (2016). https://doi.org:10.1021/acs.inorgchem.6b01941
62	Leinders, G., Bes, R., Kvashnina, K. O. & Verwerft, M. Local Structure in U(IV) and U(V) Environments: The Case of $U_3O_7$. *Inorg Chem* **59**, 4576-4587 (2020). https://doi.org:10.1021/acs.inorgchem.9b03702
63	Leinders, G. *et al.* Charge Localization and Magnetic Correlations in the Refined Structure of $U_3O_7$. *Inorganic Chemistry* **60**, 10550-10564 (2021). https://doi.org:10.1021/acs.inorgchem.1c01212
64	Nawada, H. P. *et al.* Oxidation and phase behaviour studies of the U-Ce-0 system. *Journal of Nuclear Materials* **139**, 19-26 (1986). https://doi.org:https://doi.org/10.1016/0022-3115(86)90159-5
65	Ha, Y.-K., Kim, J.-G., Park, Y.-J. & Kim, W.-H. Studies on the Air-oxidation Behavior of Uranium Dioxide I. Phase transformation from $(U_{1-y}Gd_y)O_2$ to $(U_{1-y}Gd_y)_3O_8$. *Journal of Nuclear Science and Technology* **39**, 772-775 (2002). https://doi.org:10.1080/00223131.2002.10875581
66	Kim, J.-G., Ha, Y.-K., Park, S.-D., Jee, K.-Y. & Kim, W.-H. Effect of a trivalent dopant, $Gd^{3+}$, on the oxidation of uranium dioxide. *Journal of Nuclear Materials* **297**, 327-331 (2001). https://doi.org:https://doi.org/10.1016/S0022-3115(01)00639-0
67	De Alleluia, I. B., Hoshi, M., Jocher, W. G. & Keller, C. Phase relationships for the ternary $UO_2$ $UO_2$ $REO_{1.5}$ (RE = Rr, Nd, Dy) systems. *Journal of Inorganic and Nuclear Chemistry* **43**, 1831-1834 (1981). https://doi.org:https://doi.org/10.1016/0022-1902(81)80392-2
68	McEachern, R. J., Doern, D. C. & Wood, D. D. The effect of rare-earth fission products on the rate of $U_3O_8$ formation on $UO_2$. *Journal of Nuclear Materials* **252**, 145-149 (1998). https://doi.org:https://doi.org/10.1016/S0022-3115(97)00286-9
69	Guo, X. *et al.* Energetics of a Uranothorite ($Th_{1-x}U_xSiO_4$) Solid Solution. *Chemistry of Materials* **28**, 7117-7124 (2016). https://doi.org:10.1021/acs.chemmater.6b03346
70	Shuller, L. C., Pavenayotin, N., Ewing, R. C. & Becker, U. Thermodynamic Properties of Actinide Oxide Solid Solutions. *MRS Online Proceedings Library (OPL)* **1125**, 1125-R1110-1106 (2008).
71	Evans, J. A., DeHart, M. D., Weaver, K. D. & Keiser, D. D. Burnable absorbers in nuclear reactors – A review. *Nuclear Engineering and Design* **391**, 111726 (2022). https://doi.org:https://doi.org/10.1016/j.nucengdes.2022.111726